\documentclass{article}
\usepackage{spconf,amsmath,graphicx}

\title{ImmerseDiffusion: A Generative Spatial Audio Latent Diffusion Model}

\name{Author(s) Name(s)\thanks{Thanks to XYZ agency for funding.}}
\address{Author Affiliation(s)}

\name{\hspace{-0.0in}Mojtaba Heydari $^{\dagger \ddagger }$ \hspace{2em} 
      Mehrez Souden $^{\star}$ \hspace{2em} 
      Bruno Conejo $^{\star}$ \hspace{2em} 
      Joshua Atkins $^{\star}$\thanks{$\ddagger$ Work completed during internship at Apple.}}

\address{$\hspace{-0.3in}^{\dagger}$ University of Rochester \hspace{0.5in}
      $^{\star}$ Apple \\ [0.14in] \tt \normalsize \hspace{-0.18in} mheydari@ur.rochester.edu  \hspace{0.1in}  \{msouden, bconejo, josh.atkins\}@apple.com } 

\usepackage{pifont}  
\usepackage{xcolor}  
\usepackage{float}
\usepackage{bm}
\usepackage{tikz}
\usepackage[absolute,overlay]{textpos}
\usepackage{afterpage}

\usepackage{breqn}
\usepackage{arydshln}
\begin{document}
\ninept
\maketitle

\begin{abstract}

We introduce ImmerseDiffusion, an end-to-end generative audio model that produces 3D immersive soundscapes conditioned on the spatial, temporal, and environmental conditions of sound objects. 
\textit{ImmerseDiffusion} is trained to generate first-order ambisonics (FOA) audio, which is a conventional spatial audio format comprising four channels that can be rendered to multichannel spatial output. 
The proposed generative system is composed of a spatial audio codec that maps FOA audio to latent components, a latent diffusion model trained based on various user input types, namely, text prompts, spatial, temporal and environmental acoustic parameters, and optionally a spatial audio and text encoder trained in a Contrastive Language and Audio Pretraining (CLAP) style.
We propose metrics to evaluate the quality and spatial adherence of the generated spatial audio. Finally, we assess the model performance in terms of generation quality and spatial conformance, comparing the two proposed modes: ``descriptive", which uses spatial text prompts) and ``parametric", which uses non-spatial text prompts and spatial parameters. Our evaluations demonstrate promising results that are consistent with the user conditions and reflect reliable spatial fidelity.

\end{abstract}
\section{Introduction}
\label{sec:intro}
In today's dynamic technology landscape, there is an increasing demand for immersive audio experiences across many domains, including virtual and augmented reality platforms, healthcare, education, and entertainment. Generative machine learning can play a significant role in these applications by creating audio content in dynamic and interactive ways. While generative audio models have made substantial progress recently, they are still limited to generating mono or stereo without capability to accurately pan sources to desired spatial locations 
\cite{liu2023audioldm,liu2024audioldm,evans2024fast}.

\begin{figure*}[!t]
  \centering
  \includegraphics[width=1.0\textwidth]{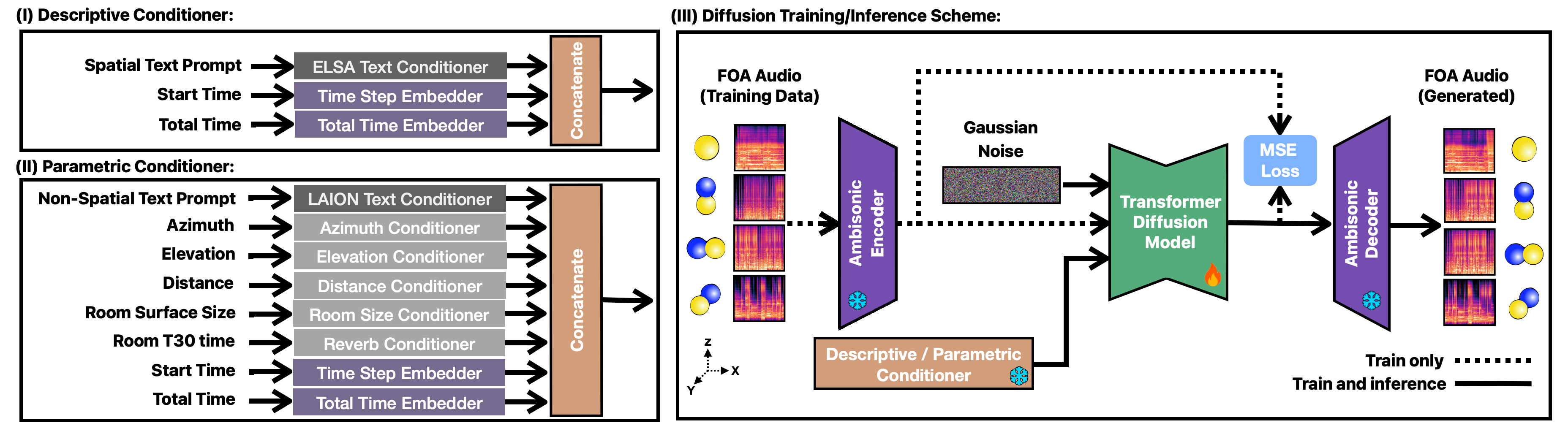}
\caption{(I) \textit{Descriptive} Conditioner: integrates ELSA~\cite{devnani2024elsa} text conditioner to encode prompts detailing the audio source and spatial and environmental context plus temporal conditioning blocks. (II) \textit{Parametric} Conditioner: including LAION~\cite{wu2023large} text encoder to provide non-spatial text embeddings along with parametric spatial, environmental, and temporal conditioning blocks. (III) \textit{ImmerseDiffusion} overall architecture, comprising the ambisonic autoencoder, conditioner, and transformer diffusion including training and inference diagrams.}

  \label{fig:scheme}
\end{figure*}

In \cite{liu2023audioldm}, Liu \textit{et al.} introduced AudioLDM which overcomes the need for large scale audio-text data annotation by using CLAP~\cite{wu2023large}. By mapping the audio to the latent domain using an audio encoder, the authors demonstrated that general audio latents can be generated then decoded to the desired final audio. In \cite{liu2024audioldm}, the authors extended AudioLDM to generate a variety of sound types including speech, music, and other sound effects. Despite its capability to generate good quality audio content, both AudioLDM variants are not designed to generate spatial audio content. 

Several other models have been proposed to produce stereo audio. MusicGen ~\cite{copet2024simple} employs a single-stage transformer language model combined with efficient token interleaving patterns to create stereo-channel music conditioned on textual descriptions or melodic features. Jen-1~\cite{li2023jen} uses a transformer latent diffusion structure to generate text-conditioned stereo-channel music. Moûsai~\cite{schneider2023mo} uses a two-stage cascade latent diffusion model with a 1D convolutional network-based U-Net structure to generate stereo music. Stable Audio~\cite{evans2024fast} adopts a 1D convolutional network with temporal conditioning in a U-Net configuration to produce stereo music. Its subsequent version \cite{evans2024long} incorporates a transformer-based diffusion model with an enhanced codec to generate extended stereo-channel music audio. Although these stereo models can generate two-channel audio, they lack the ability to map the audio sources to desired spatial locations. For example, we found that they are unable to generate sound exclusively from the left channel in response to a prompt similar to \textit{''a dog barking on the left''}. This limitation hinders their appropriateness in creating stereo audio experiences with precise spatial rendering.

This paper introduces \textit{ImmerseDiffusion}, a text-conditioned generative spatial audio model designed to produce 3D Ambisonic soundscapes with localization in horizontal, vertical, and distance dimensions, while also accounting for temporal and environmental factors like room size and reverberation. Our model generates FOA-domain audio, and features two operational modes: \textit{descriptive} and \textit{parametric}. The \textit{descriptive} model generates spatial audio based on textual descriptions of sound sources and their spatial and environmental details, suited for narrative-driven applications such as cinematic audio. The \textit{parametric} model combines textual descriptions of sound sources with numerical spatial parameters, ideal for machine-centered uses like game engines and virtual simulations. To the best of our knowledge, this is the first spatial audio generative model. Consequently, we introduce new metrics to evaluate the quality of the generated FOA audio. These metrics include Ambisonics Fréchet Audio Distance (FAD), spatial Kullback–Leibler (KL) divergence, and spatial CLAP scores. Additionally, we assess the spatial accuracy using azimuth, elevation, and distance L1 scores, besides spatial angle which are all based on sound intensity vectors \cite{intensityvector}.
 
\section{Methodology}
\label{sec:related_works}
\textit{ImmerseDiffusion} integrates three main components: a spatial autoencoder designed to encode 4-channel ambisonic domain signals into a continuous latent domain and decode back to the waveform representation, a cross-attention-based conditioner block that incorporates spatial information in two distinct manners tailored to \textit{descriptive} and \textit{parametric} generation modes, 
and a transformer-based diffusion model operating on the autoencoder latents. This architecture enables the generation of spatial audio aligned with audio source descriptions, spatial, environmental and temporal cues, enhancing realism and immersion across various interactive and narrative-driven applications.

\subsection{Spatial Audio Format}

The Ambisonic surround-sound system encodes and reproduces sound directions and amplitudes to deliver immersive 3D audio experiences. For simplicity and considering the fact that the first-order Ambisonics (FOA) format generally provides adequate spatial resolution~\cite{malham19953}, this paper focuses exclusively on generating FOA components. However, it can be generalized to higher-order Ambisonics for improved spatial fidelity, albeit with increased complexity. FOA achieves spatial rendering using four-channel components, denoted as W, X, Y, and Z as follows:
\begin{equation}
\begin{aligned}
W &= \frac{1}{\sqrt{2}} p, & \quad X &= p \cos(\theta) \cos(\phi) \\
Y &= p \sin(\theta) \cos(\phi), & \quad Z &= p \sin(\phi) 
\end{aligned}
\end{equation}
where \( p \) is sound pressure, \( W \) is the omnidirectional component, \( X \) is the front-back directional component, defined by the azimuth angle \(\theta\)$ \;\in (-\pi, \pi]$,  and elevation angle \(\phi\)$\;\in (-\pi/2, \pi/2]$, \( Y \) is the left-right directional component, defined by the azimuth angle \(\theta\) and elevation angle \(\phi\), and \( Z \) is the up-down directional component, defined by the elevation angle \(\phi\).
Both the training data and the generated audio will conform to this standard.
\subsection{Spatial Codec}
\label{ssec:codec}
Large model sizes, high-dimensional audio data, iterative diffusion processes, and the need to capture long-term structures necessitate efficient computation in generative audio models. Autoencoder architectures are commonly employed to manage these demands by operating within compressed latent domains. For our model, which generates four FOA components simultaneously, compression is even more critical.

To address compression, various approaches have been explored in the literature. Some models leverage autoencoders with quantized latents such as Vector Quantized Variational Autoencoders VQ-VAE (e.g.,~\cite{dhariwal2020jukebox})  
or Residual Vector Quantization RVQ-VAE (e.g.,~\cite{borsos2023audiolm}). 
Other models leverage continuous VAE latents to avoid quantization information loss~\cite{liu2024audioldm,evans2024fast,evans2024long}.
Following \cite{evans2024fast}, we utilize the 1D convolutional U-Net autoencoder from the Descript Audio Codec (DAC)~\cite{kumar2024high}
and replace the discrete RVQ bottleneck with a continuous VAE bottleneck. We also eliminate the decoder's $tanh()$ activation to prevent harmonic distortion, as noted in \cite{evans2024long}. Additionally, we modify the training loss functions to suit the FOA components. Unlike stereo codecs, which leverage losses like Mid-Side Short Time Fourier Transform (MS-STFT) for proper stereo reconstruction, FOA does not need such losses due to the inherent orthogonality of the $X$, $Y$, and $Z$ channels. The training loss function of our FOA codec is as follows:
\begin{dmath}
\mathcal{L}_{\text{codec}} = \frac{\lambda_{\text{mrstft}}}{4} \mathcal{L}_{\text{mrstft\_W}} 
+ \frac{\lambda_{\text{mrstft}}}{4} \mathcal{L}_{\text{mrstft\_X}} 
+ \frac{\lambda_{\text{mrstft}}}{4} \mathcal{L}_{\text{mrstft\_Y}} 
+ \frac{\lambda_{\text{mrstft}}}{4} \mathcal{L}_{\text{mrstft\_Z}} 
+ \lambda_{\text{kl}} \mathcal{L}_{\text{kl}} 
+ \lambda_{\text{adv}} \mathcal{L}_{\text{adv}} 
+ \lambda_{\text{fm}} \mathcal{L}_{\text{fm}}
\end{dmath}
\label{eq:codec_loss}

In this context, $\mathcal{L}_{\text{mrstft}}$ represents the Multi-Resolution STFT loss for each channel, collectively contributing to the generation loss. $\mathcal{L}_{\text{kl}}$ denotes KL divergence loss applied to the VAE bottleneck, $\mathcal{L}_{\text{adv}}$ signifies the adversarial (discrimination) loss, while $\mathcal{L}_{\text{fm}}$ refers to the feature matching loss. The $\lambda$ parameters are weights perceptually assigned to adjust each corresponding loss contribution in total loss~\cite{evans2024fast}. 
Our ambisonic codec model transforms a 4-channel FOA signal with length $L$ into $64$ channel latent with length $L/2048$ with an overall compression factor of $128$.

\begin{table*}[ht]
\centering
\caption{Specifications and Quality and Spatial Accuracy Evaluation Results of the trained FOA Codecs on Spatial AudioCaps.}
\label{tab:codec_structure}
\begin{tabular}{l:cc:cc:cccccccc} 
\hline
\textbf{FOA Codec Model} & 
\textbf{\begin{tabular}[c]{@{}c@{}}Downsampling \\ Rate\end{tabular}} & 
\textbf{\begin{tabular}[c]{@{}c@{}}Latent \\ Size\end{tabular}} & 
\textbf{\begin{tabular}[c]{@{}c@{}}STFT \\ Distance\end{tabular}} & 
\textbf{\begin{tabular}[c]{@{}c@{}}Mel \\ Distance\end{tabular}} & 
\textbf{L1\(_{\text{azimuth}}\)}& 
\textbf{L1\(_{\text{elevation}}\)} & 
\textbf{L1\(_{\text{distance}}\)} & 
\textbf{Angular\(_{\text{difference}}\)} \\ 
\hline

\textbf{1D-Conv U-Net(32X)} & 
1024 & 
128 & 
1.69 & 
2.83 & 
1.55 & 
0.29 & 
2.05 & 
1.56 \\ 

\textbf{1D-Conv U-Net (64X)} & 
1024 & 
64 & 
1.70 & 
2.62 & 
1.57 & 
0.24 & 
1.82 & 
1.57 \\ 

\textbf{1D-Conv U-Net (128X)} & 
2048 & 
64 & 
1.73 & 
2.55 & 
1.57 & 
0.16 & 
1.49 & 
1.57 \\ 

\textbf{1D-Conv U-Net-$\beta$ (128X)} & 
2048 & 
64 & 
2.33 & 
4.13 & 
1.43 & 
0.15 & 
1.53  & 
1.51 \\ 
\hline

\textbf{Measurement Error} & 
--- & 
--- & 
--- & 
--- & 
0.17 & 
0.14 & 
1.36   & 
1.03 \\ 
\hline
\end{tabular}
\begin{tikzpicture}[overlay, remember picture]
    \node[anchor=west] at (-0.05, 2.3) {\scalebox{1.3}{\(\downarrow\)}};
    \node[anchor=west] at (1.51, 2.3) {\scalebox{1.3}{\(\downarrow\)}};
    \node[anchor=west] at (3.1, 2.3) {\scalebox{1.3}{\(\downarrow\)}};
    \node[anchor=west] at (4.7, 2.3) {\scalebox{1.3}{\(\downarrow\)}};
    \node[anchor=west] at (6.25, 2.3) {\scalebox{1.3}{\(\downarrow\)}};
    \node[anchor=west] at (8.7, 2.3) {\scalebox{1.3}{\(\downarrow\)}};
\end{tikzpicture}
\label{tab:codec}
\end{table*}

\subsection{Spatial, Temporal and Environmental Conditioning}
\label{ssec:descriptive-cond}

Many generative audio methods use text prompts for user control~\cite{dhariwal2020jukebox, borsos2023audiolm}. Recent approaches have transitioned from pure language models to CLAP-based models, such as LAION CLAP~\cite{wu2023large} for audio and MuLan~\cite{huang2022mulan} for music, due to their enhanced performance~\cite{evans2024fast,liu2024audioldm,zhu2023edmsound}. These models are effective in aligning language and audio embeddings, enabling cross-modality training and inference~\cite{liu2024audioldm}. Additionally, various conditioning types are explored, including control over temporal order, pitch, energy~\cite{guo2024audio}, music genre~\cite{dhariwal2020jukebox}, 
and rhythm and chords~\cite{lan2024musicongen}.

\textit{ImmerseDiffusion}  offers spatial control in audio generation via two models: \textit{descriptive} and \textit{parametric}. The \textit{descriptive} model utilizes text-based prompts to define the sound source, its spatial attributes, and environmental acoustics. In contrast, the \textit{parametric} model uses non-spatial text prompts only for sound source descriptions while providing numerical values for spatial and environmental parameters.  Moreover, in line with previous stereo diffusion models ~\cite{evans2024fast,evans2024long}, both models employ temporal conditioning, including start time and total time as well. In Figure~\ref{fig:scheme}, (I) and (II) illustrate how \textit{descriptive} and \textit{parametric} conditioning incorporate location and environmental parameters through text or a combination of text and numerical values, along with temporal conditioning.

To generate conditioning spatial text embeddings for the \textit{descriptive} model, we use the Embeddings for Language and Spatial Audio (ELSA) \cite{devnani2024elsa} model's text encoder. Trained on spatial audio-text pairs using synthetic spatial variants of AudioCaps~\cite{kim2019audiocaps}, FreeSound~\cite{font2013freesound}, and Clotho~\cite{drossos2020clotho}, ELSA demonstrated superior performance in spatial audio retrieval tasks~\cite{devnani2024elsa}.
ELSA's design for comprehending spatial qualities makes it ideal for text conditioning in our model. Thanks to ELSA text conditioner and the mentioned spatial training datasets, our proposed \textit{descriptive} model considers spatial description details (e.g., \textit{right corner, far left, front top}) and environment acoustics (e.g., \textit{medium-sized acoustically damped room, large highly reverberant hall}) in spatial audio generation. In addition to that, two cross-attention conditioners are employed to include the mentioned temporal conditions (i.e. start time and total time). 

For the \textit{parametric} model, we replace ELSA text encoder with the LAION CLAP~\cite{wu2023large} text encoder, which is solely trained on non-spatial prompts to validate \textit{parametric} conditioning and avoid potential spatial bias from ELSA embeddings. Alongside text embeddings, this model includes three numerical spatial conditioning parameters (azimuth, elevation, distance) and two environmental acoustic conditioning parameters (room size, 30dB decay time). It also comprises two temporal conditioning parameters, namely, start time and total time, similar to the \textit{descriptive} model.

\subsection{Diffusion Model}
\label{ssec:diffusion}
Several generative audio latent diffusion models utilize 1D convolutional U-Net structures for their diffusion blocks (e.g.,~\cite{liu2024audioldm, evans2024fast}). However, the Diffusion Transformer (DiT), initially proposed for image generation~\cite{peebles2023scalable}, has recently been adapted for music generation~\cite{evans2024long, levy2023controllable}. In our work, we employ a DiT that features a cascade of blocks that include self-attention, cross-attention, and gated MLP components with layer normalization and skip connections, similar to that used in~\cite{evans2024long}, with structural adoptions for FOA-domain. We adjust conditioning token dimensions to $512$ and introduce a projecting conditioning layer to meet dimensionality requirements. The model is trained using the v-objective approach~\cite{salimans2022progressive} to minimize the Mean Squared Error (MSE) between the true \( v \) and the predicted \( v_\theta(z_t, t, c) \) at timestep \( t \), conditioned on \( c \) and given the latent variable \( z_t \) as follows:
\begin{equation}
\mathcal{L}_{\text{Diffusion}} = \left\|v_\theta(z_t, t, c) - v\right\|_2^2
\end{equation}

\subsection{Datasets}
\label{ssec:datasets}
We trained the ambisonic codec and both \textit{descriptive} and \textit{parametric} \textit{ImmerseDiffusion}  models using synthetically spatialized versions of the FreeSound~\cite{font2013freesound}, AudioCaps~\cite{kim2019audiocaps}, and Clotho~\cite{drossos2020clotho} datasets, as provided in~\cite{devnani2024elsa}. To improve generalization and performance on speech audio, we also included a spatialized version of LibriSpeech~\cite{panayotov2015librispeech} and the corresponding FOA noise dataset, detailed in~\cite{sarabia2023spatial}. Each dataset contains spatialized audio, original captions, spatial and environmental parameters, and spatially derived captions, as described in~\cite{devnani2024elsa}. Datasets information is provided in Table~\ref{tab:datasets}.
\begin{table}[ht]
\centering
\caption{Datasets used for training and evaluation of the models}
\label{tab:datasets}
\begin{tabular}{lcc}
\hline
\textbf{Datasets}                  & \textbf{\begin{tabular}[c]{@{}c@{}}Number of \\ Samples\end{tabular}} & \textbf{\begin{tabular}[c]{@{}c@{}}Duration\\ (hrs)\end{tabular}} \\ \hline

\textbf{Spatial Clotho~\cite{devnani2024elsa,drossos2020clotho}}         & 8,546                                                                & 55.0                                                            \\ \hline
\textbf{Spatial AudioCaps~\cite{devnani2024elsa,kim2019audiocaps}}         & 98,459                                                                & 258.12                                                            \\ \hline
\textbf{Spatial Freesound~\cite{devnani2024elsa,font2013freesound}}         & 783,033                                                               & 4,425.53                                                          \\ \hline
\textbf{Spatial LibriSpeech~\cite{sarabia2023spatial,panayotov2015librispeech}}       & 132,377                                                               & 658.6                                                             \\ \hline
\textbf{Spatial LibriSpeech Noise~\cite{sarabia2023spatial,panayotov2015librispeech}} & 132,377                                                               & 658.6                                                              \\ \hline
\end{tabular}
\end{table}

\subsection{Training}
\label{ssec:training}
We train our models with a cluster of A100-80GB GPUs.
For the ambisonic audio codec, we use a batch size of 96 and an excerpt length of 1.48 seconds for 600K steps. 
The spatial codec generator and discriminator were trained with base learning rates of $10^{-4}$ and $2. \times 10^{-4}$ respectively, using the AdamW optimizer, applying a $10^{-3}$ decay factor and momentum parameters of 0.8 and 0.99.

Both \textit{descriptive} and \textit{parametric} models share the same training setup. We trained the Diffusion transformers with frozen ambisonic codec and conditioner weights with a batch size of 576 and a context window of 5.95 seconds for 500K steps. The base learning rate was set to $10^{-4}$ with the AdamW optimizer, employing momentum values of 0.9 and 0.99. Spatial and non-spatial text conditioning for \textit{descriptive} and \textit{parametric} models utilized the official pre-trained ELSA~\cite{devnani2024elsa} and LAION CLAP~\cite{wu2023large} models, respectively.

\begin{table*}[ht]
\centering
\caption{Quality and Spatial Accuracy Evaluation of \textit{ImmerseDiffusion}  for both \textit{descriptive} and \textit{parametric} Models on Spatial AudioCaps.} 
\label{tab:diffusion}
\renewcommand{\arraystretch}{1.2}
\begin{tabular}{l:ccc:cccc}
\hline
\textbf{Diffusion Model} & \textbf{FAD}\textnormal{\textsubscript{ELSA}} & \textbf{KL}\textnormal{\textsubscript{ELSA}} & \hspace{0.5em}\textbf{CLAP}\textnormal{\textsubscript{ELSA}\hspace{0.4em}} & 
\hspace{1.1em}\textbf{L1}\textnormal{\textsubscript{\(\theta\)\phantom{xxx}}} & 
\hspace{1.1em}\textbf{L1}\textnormal{\textsubscript{\(\phi\)\phantom{xxx}}} & 
\hspace{1.1em}\textbf{L1}\textnormal{\textsubscript{d}\phantom{xxx}} & 
\textbf{$\Delta$}\textnormal{\textsubscript{Spatial-Angle}} \\ 
\hline
\textbf{Descriptive\phantom{Wider}} & 0.28 & 0.11 & 0.64 & 1.08 & 0.43 & 1.88m & 1.35\\    
\textbf{Parametric\phantom{Wider}} & 1.75 & 0.094 & 0.59 & 0.36 & 0.34 & 1.92m & 1.12 \\  
\hline
\end{tabular}

\begin{tikzpicture}[overlay, remember picture]
    \node[anchor=west] at (-3.3, 1.18) {\scalebox{1.3}{\(\downarrow\)}};
    \node[anchor=west] at (-1.86, 1.18) {\scalebox{1.3}{\(\downarrow\)}};
    \node[anchor=west] at (0.05, 1.18) {\scalebox{1.3}{\(\uparrow\)}};
    \node[anchor=west] at (1.49, 1.18) {\scalebox{1.3}{\(\downarrow\)}};
    \node[anchor=west] at (3.15, 1.18) {\scalebox{1.3}{\(\downarrow\)}};
    \node[anchor=west] at (4.75, 1.18) {\scalebox{1.3}{\(\downarrow\)}};
    \node[anchor=west] at (7.10, 1.18) {\scalebox{1.3}{\(\downarrow\)}};
\end{tikzpicture}

\end{table*}

\subsection{Evaluation Metrics}
\label{ssec:metrics}
The quality of FOA codec models is evaluated using STFT and MEL distances between the original and reconstructed FOA audio from the test set using AuraLoss~\cite{steinmetz2020auraloss} with default settings similar to~\cite{evans2024fast, evans2024long}.

We evaluate the plausibility of generated FOA audio clips by computing FAD which compares the statistical properties of the generated and reference audio embeddings. To compute the reference and generated FOA audio embeddings,  we used ELSA audio encoder that supports FOA audio at 48 kHz. Additionally, we report KL divergence using a pre-trained ELSA.

The CLAP score is computed as the cosine similarity between the conditioning spatial text embeddings and the generated FOA audio embeddings, using the ELSA model. For the \textit{parametric} model, KL divergence and CLAP score are computed using corresponding spatial captions from the test set, despite the model being trained on non-spatial captions and spatial parameters.   

To measure the spatial accuracy of FOA audio generated by \textit{ImmerseDiffusion}, we report metrics based on ground truth and estimated azimuth (\(\theta\)), elevation (\(\phi\)), and distance (d) using equation set (\ref{eq:combined}). The intensity vectors $I_x$, $I_y$ and $I_z$ are derived by multiplying the omnidirectional channel with the corresponding directional channels, as outlined in (\ref{eq:intensity}).

\begin{equation}
\resizebox{0.3\textwidth}{!}{$
I_x = W \cdot X, \quad \quad I_y = W \cdot Y, \quad \quad I_z = W \cdot Z
$}
\label{eq:intensity}
\end{equation}
\begin{equation}
\resizebox{\columnwidth}{!}{$
\begin{aligned}
\text{\(\theta\)} = \tan^{-1}\left(\frac{I_y}{I_x}\right), \quad
\text{\(\phi\)} = \tan^{-1}\left(\frac{I_z}{\sqrt{I_x^2 + I_y^2}}\right), \quad
\text{d} = \sqrt{I_x^2 + I_y^2 + I_z^2}
\end{aligned}
$}
\label{eq:combined}
\end{equation}

For all mentioned three parameters, we report the L1 norm between the ground truth and estimated values for azimuth, elevation and distance, across the test set. To calculate the L1 norm for azimuth, we use the circular difference method (see Equation \ref{eq:circular_difference}). This method accounts for the circular nature of azimuth angles, ensuring we measure the shortest angular distance between the ground truth  \(\theta\) and the estimated azimuth $\hat{\theta}$. It avoids issues with linear differences, where angles like -3.13 and +3.13 are in nearly the same direction but appear far apart in a linear measurement.

\begin{equation}
\text{L1}_{\text{\(\theta\)}} = \|\min \left( \left| \text{\(\theta\)} - \hat{\theta} \right|, 2\pi - \left| \text{\(\theta\)} - \hat{\theta} \right| \right)||_1
\label{eq:circular_difference}
\end{equation}
To quantify the difference between the ground truth and estimated direction of arrival (DoA), we also include their spatial angle ${\Delta}_{\text{Angular}}$~\cite{van2012heavenly} calculated using equations~(\ref{eq:alpha}) and ~(\ref{eq:angular_distance}) as follows:
\begin{equation}
a = \sin^2\left(\frac{{\Delta}_{\phi}}{2}\right) + \cos(\phi) \cdot \cos({\hat{\phi}}) \cdot \sin^2\left(\frac{{\Delta}_{\theta}}{2}\right)
\label{eq:alpha}
\end{equation}
\begin{equation}
{\Delta}_{\text{Spatial-Angle}} = 2 \cdot \arctan2\left(\sqrt{a}, \sqrt{1 - a}\right)
\label{eq:angular_distance}
\end{equation}
where ${\Delta}_{\phi}$ and ${\Delta}_{\theta}$ denote the linear and circular differences for elevation and azimuth respectively, and ${\phi}$ and $\hat{\phi}$ represent the ground truth and estimated elevations.

\section{Results and Discussions}
\label{sec:results}
We summarize the evaluation results of the ambisonic codec models in Table~\ref{tab:codec}. All models were trained with identical hyper parameters. The 1D-Conv U-Net (32X) is based on the baseline~\cite{evans2024fast} structure, with the $tanh$ activation removed. The 1D-Conv U-Net (64X) and 1D-Conv U-Net (128X) follow the same structure with 2X and 4X higher compression ratios, respectively. We also include the 1D-Conv U-Net-$\beta$ (128X) builds on~\cite{evans2024long} baseline, to evaluate incorporating a learned snake activation frequency $\beta$. This model maintains the same compression ratio as the 1D-Conv U-Net (128X), which is twice that of the original model~\cite{evans2024long}.

Comparing the first three models shows that higher compression achieves similar FOA reconstruction quality, indicating that increased compression is suitable for this task. Notably, as compression increases, MEL performance improves, likely because the model focuses more on reconstructing lower frequencies, which are crucial for MEL distance. Additionally, comparing 1D-Conv U-Net (128X) with 1D-Conv U-Net-$\beta$ (128X) reveals that the model without the learned $\beta$ outperforms the one with learned $\beta$. The inconsistent efficacy of the learned $\beta$ is also noted in previous work on music codecs~\cite{evans2024long}. 

Table~\ref{tab:diffusion} shows the quality and spatial accuracy evaluations for both \textit{descriptive} and \textit{parametric} \textit{ImmerseDiffusion}  models.
The \textit{descriptive} model significantly outperforms the \textit{parametric} model in FAD\textsubscript{ELSA} score. However, this advantage may be influenced by the use of ELSA embeddings as a conditioner in the \textit{descriptive} model, which the \textit{parametric} model lacks. 
In terms of KL divergence and CLAP scores, on the other hand, the \textit{parametric} model performs similarly to the \textit{descriptive} model, slightly outperforming in one and under-performing in the other. This comparable performance underlines the effectiveness of the \textit{parametric} training. In fact, even though the \textit{parametric} model is trained on non-spatial prompts, it is capable of producing spatially plausible audio content thanks to the additional spatial parametrization.

We report the spatial accuracy of both models using the metrics proposed in Section \ref{ssec:metrics}. Given that localization estimation inherently includes some error, we also account for measurement error, which is the lower bound on the error that the model can achieve, by comparing the estimated spatial parameters with the ground truth labels. Both models demonstrate high localization accuracy when considering these errors. However, the \textit{parametric} model significantly outperforms the \textit{descriptive} model across all Direction of Arrival (DoA) metrics. This superior DoA performance is expected, as the \textit{parametric} model uses precise absolute location values, offering more accurate spatial conditioning compared to the broader ranges in the \textit{descriptive} labels. In contrast, for distance estimation, both models perform similarly.

It is worth mentioning that the signal processing methods used for spatial adherence evaluation offer two key advantages: simplicity, which is a crucial criterion for evaluation metrics, and generality, as they do not require pre-trained models or pose the risk of training data biases, unlike deep learning methods. However, due to higher lower-bound errors in some metrics (e.g., distance or spatial angle differences) as shown in Table~\ref{tab:diffusion}, future extension of this work can include leveraging deep learning-based methods such as ~\cite{adavanne2018sound} to improve measurement accuracy.

\section{Conclusion}
\label{sec:conclusion}

This paper introduced ImmerseDiffusion, an end-to-end model for generating 3D immersive soundscapes based on spatial, temporal, and environmental conditions. Using a spatial audio codec and a latent diffusion model, \textit{ImmerseDiffusion} generates first-order ambisonics (FOA) audio. We presented two models: a \textit{descriptive} model that creates spatial audio from text prompts detailing sound sources and contexts, and a \textit{parametric} model that uses text prompts with explicit spatial and environmental parameters. Experimental evaluations demonstrated that \textit{ImmerseDiffusion} produces spatially accurate audio according to user-defined conditions.

\bibliographystyle{IEEEbib}
\bibliography{ImmerDiffusion}

\end{document}